\documentclass[aps,pra,reprint,superscriptaddress,nofootinbib]{revtex4-2}

\usepackage[utf8]{inputenc}
\usepackage[T1]{fontenc}
\usepackage{lmodern}   

\usepackage{graphicx}
\usepackage{amsmath,amssymb,bm}
\usepackage{siunitx}
\usepackage{hyperref}
\usepackage{braket}
\usepackage{physics}
\usepackage{xcolor}

\usepackage{orcidlink}

\hypersetup{
  colorlinks=true,
  linkcolor=blue,
  citecolor=blue,
  urlcolor=blue
}

\begin{document}

\title{Phase Independent Measurement of Weak Coherent Optical Signals}


\author{Lani Chastain\orcidlink{0009-0005-6101-9367}}
\thanks{L.C. and P.D. contributed equally to this work.}
\affiliation{Department of Physics \& Astronomy, University of Tennessee, Chattanooga, TN 37403, USA}

\author{Priya Drashni\orcidlink{0009-0009-5676-6033}}
\thanks{L.C. and P.D. contributed equally to this work.}
\affiliation{Department of Electrical \& Computer Engineering, FAMU–FSU College of Engineering, Florida State University, Tallahassee, FL 32306, USA}

\author{Mahadeva Chanda Durjoy\orcidlink{0009-0008-2566-6497}}
\affiliation{Institute for Quantum Science and Engineering, Department of Physics \& Astronomy, Texas A\&M University, College Station, TX 77843, USA}

\author{Hari P.~Lamsal}
\affiliation{Department of Physics \& Astronomy, University of Tennessee, Chattanooga, TN 37403, USA}

\author{Girish S. Agarwal\orcidlink{0000-0002-8320-4708}}
\email{girish.agarwal@ag.tamu.edu}
\affiliation{Institute for Quantum Science and Engineering, Department of Physics \& Astronomy, Texas A\&M University, College Station, TX 77843, USA}
\affiliation{Department of Biological and Agricultural Engineering, Texas A\&M University, College Station, TX 77843, USA}

\author{Tian Li\orcidlink{0000-0003-2993-0386}}
\email{tli6@fsu.edu}
\affiliation{Department of Electrical \& Computer Engineering, FAMU–FSU College of Engineering, Florida State University, Tallahassee, FL 32306, USA}


\begin{abstract}
We develop a quantum sensing framework for the phase-independent detection of weak coherent optical displacements based on SU(1,1) interferometry. Unlike conventional quantum measurement protocols that require prior knowledge of the signal phase and coherent homodyne detection, the proposed approach estimates the displacement magnitude independently of its phase. We show that, under ideal lossless conditions, a conventional SU(1,1) interferometer employing only total intensity detection saturates the quantum Cramér–Rao bound for displacement magnitude estimation. We further derive the analytical expression of the quantum Cramér–Rao bound and the sensitivity of the conventional SU(1,1) interferometer with total intensity detection and systematically investigate its performance in the presence of optical loss. The proposed phase-independent intensity detection scheme achieves comparable performance over experimentally relevant operating regimes while eliminating the need for local oscillators, phase locking, and quadrature tracking. These results establish SU(1,1)-based intensity detection as a practical platform for phase-independent quantum sensing.
\end{abstract}



\maketitle

\section{\label{sec:introduction}Introduction}




The detection of weak optical signals with sensitivity beyond classical limits is a fundamental task in precision measurement. By exploiting nonclassical states of light, such as squeezed and entangled states, quantum-enhanced metrological techniques can surpass the performance achievable with classical optical fields. These quantum advantages have enabled significant advances in a variety of applications, including gravitational-wave detection with Advanced LIGO and Virgo~\cite{LIGO_2019,VIRGO_2019,LIGO_2013}, dark-matter axion searches~\cite{Backes_2021,Zheng_2016}, quantum-enhanced imaging~\cite{He_2023,Taylor_2013,Lopaeva_2013,Gregory_2020,Blakey_2022}, and continuous-variable quantum information processing~\cite{Furusawa_1998,Yokoyama_2013}.

Among quantum-enhanced sensing architectures, the SU(1,1) interferometer, introduced by Yurke \emph{et al.}~\cite{Yurke_1986}, has emerged as a powerful platform for precision metrology due to its ability to surpass the standard quantum limit (SQL) and, in principle, approach the Heisenberg limit of $1/\bar{N}$, where $\bar{N}$ is the average photon number in the interferometer. The SU(1,1) interferometer replaces the passive beam splitters of a conventional Mach--Zehnder interferometer with active nonlinear elements, such as optical parametric amplifiers (OPAs) or four-wave mixers, and exploits coherent quantum amplification and de-amplification to enhance phase sensitivity while minimizing added noise~\cite{Caves_2020,Ou_2012}.

Most applications of nonlinear SU(1,1) interferometry have focused on phase estimation, where the optimal measurement strategy is derived under the assumption that the relevant field quadrature is known and can be phase matched to a local oscillator~\cite{Anderson_2017,D_Li_2014,Ou_2012,Ou_2020}. In many practical scenarios, however, the signal phase is not known \emph{a priori}, may fluctuate during the measurement process, or is fundamentally inaccessible. Under such conditions, the sensing task transitions from single-parameter phase estimation to the detection of a weak displacement with an unknown direction in phase space~\cite{Genoni_2013,Hanamura_2023}. We refer to this operating regime as \emph{phase-independent}, wherein the performance of the detection scheme depends solely on the magnitude of the displacement and is independent of the associated phase information.

Unknown-phase displacement sensing is intrinsically a two-parameter problem that can be addressed by engineering an appropriate measurement scheme for a SU(1,1)-based interferometer. The SU(1,1) configuration can be viewed generally as a “displacement detector” constructed as a sequence of two-mode squeezing, signal injection, and reverse squeezing, with the relevant signal information carried by joint EPR-type quadrature observables rather than by a single phase-matched fringe signal~\cite{Caves_2020}. This reformulation connects SU(1,1) interferometry directly to the broader framework of multiparametric displacement estimation and motivates re-examining its performance when the signal phase is unknown or uncontrolled, as in recent demonstrations of sub-classical joint displacement sensing~\cite{Hanamura_2023}.

In this work, we investigate a phase-independent scheme for detecting weak optical displacements based on a SU(1,1) interferometric architecture. The proposed approach relies \textit{solely on intensity measurements}, thereby eliminating the need for coherent homodyne detection. Unlike conventional quantum sensing protocols that assume prior knowledge of the signal phase and optimize sensitivity along a selected quadrature, we consider the more general scenario in which the displacement phase is unknown. Working in the Heisenberg picture, we analyze how two-mode squeezing generated within a SU(1,1) interferometer distributes the displacement information across signal and idler modes, enabling a sensitivity to the displacement magnitude that is independent of the signal phase. Using the quantum Fisher information, we derive the fundamental bound on the estimation of displacement magnitude. We then derive analytical sensitivity expressions for direct intensity detection. The effects of propagation and detection loss are incorporated through a unified beam-splitter model, allowing a systematic comparison of robustness and performance. The analytical results presented in this article were derived using a custom program implemented with the \textit{Mathematica} package \textit{NCAlgebra}~\cite{NCAlgebra}.


\section{\label{sec:theoretical_framework}Theoretical Results}
\subsection{\label{sec:ideal_model} Ideal Model without Loss}

\begin{figure}[t]
\centering
\includegraphics[width=\columnwidth]{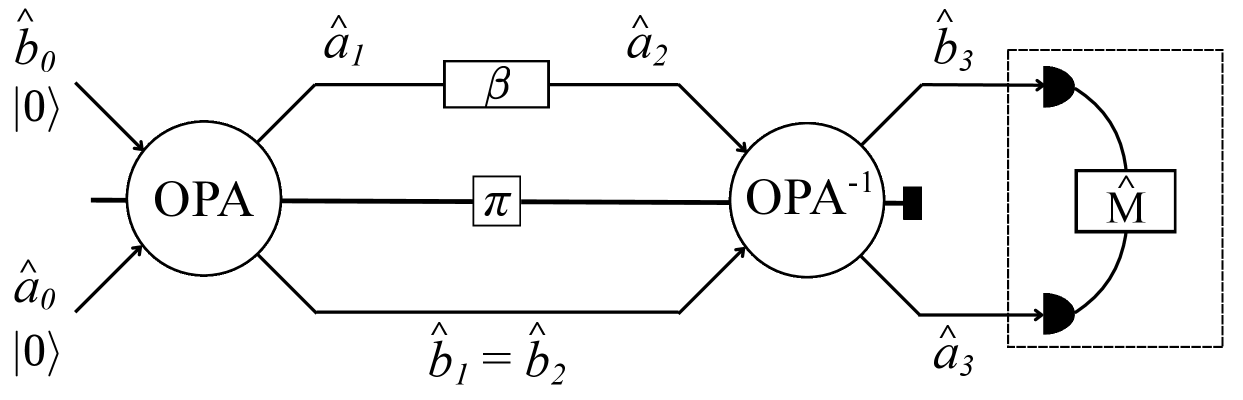}
\caption{Schematic of our proposed total intensity measurement scheme based on an idealized lossless SU(1,1) interferometer. Input modes $\hat{a}_{0}$ and $\hat{b}_{0}$ are injected into the first OPA, generating correlated output modes $\hat{a}_{1}$ and $\hat{b}_{1}$. A weak coherent displacement operation $\hat{D}(\beta)$ is subsequently applied to mode $\hat{a}_{1}$ before the two modes are recombined in a second OPA. The central dark line denotes a strong classical pump field, which acquires a $\pi$ phase shift prior to driving the second OPA, thereby realizing the inverse parametric transformation. The measurement operator $\hat{M}$ is performed by detecting the total output intensity of modes $\hat{a}_{3}$ and $\hat{b}_{3}$.}
\label{fig:toy_model}
\end{figure}

We first consider the ideal lossless SU(1,1) interferometric configuration shown in Fig.~\ref{fig:toy_model}. In the Heisenberg picture, the action of the first optical parametric amplifier (OPA) on the two input modes, $\hat{a}_0$ and $\hat{b}_0$, is described by the two-mode Bogoliubov transformation~\cite{Yurke_1986,gerry2005introductory}:
\begin{equation}
\begin{split}
    \hat{a}_1 &= \hat{a}_0 \cosh r + \hat{b}_0^\dagger\sinh r, \\
    \hat{b}_1 &= \hat{b}_0\cosh r+\hat{a}_0^\dagger\sinh r,
\label{eqs:OPA1}
\end{split}
\end{equation}
where, without loss of generality, the pump phase has been chosen to be zero, and $r$ is the two-mode squeezing parameter. A weak coherent displacement operation, described by the operator $\hat{D}(\beta)=\exp(\beta\hat{a}_1^\dagger-\beta^*\hat{a}_1)$, is applied to mode $\hat{a}_1$, where $\beta=|\beta|e^{i\theta}$ represents the complex displacement to be measured. Here, $|\beta|$ and $\theta$ denote the displacement amplitude and phase, respectively. The displaced mode is then given by
\begin{equation}
    \hat{a}_2 = \hat{D}^\dagger(\beta)\,\hat{a}_1\,\hat{D}(\beta) = \hat{a}_1 + \beta,
    \label{eq:displacement}
\end{equation}
which follows directly from the Baker--Campbell--Hausdorff lemma,
\begin{equation}
    e^{i\hat{G}\lambda}\hat{A}e^{-i\hat{G}\lambda} = \hat{A}+i\lambda[\hat{G},\hat{A}]+\frac{(i\lambda)^2}{2!}[\hat{G},[\hat{G}, \hat{A}]] + ...,
\end{equation}
together with $[\beta \hat{a}_1^\dagger - \beta^* \hat{a}_1,\, \hat{a}_1] = -\beta$ and all vanishing higher-order commutators. As the weak displacement operator acts exclusively on the $\hat{a}_1$ mode, the $\hat{b}_2$ mode remains unchanged with $\hat{b}_2=\hat{b}_1$.

Driven by the same pump field with a relative $\pi$-phase shift, the second OPA implements the inverse two-mode Bogoliubov transformation, corresponding to the inverse squeezing operation with squeezing parameter $-r$:
\begin{equation}
\begin{split}
    \hat{a}_3 &= \hat{a}_2 \cosh r - \hat{b}_2^\dagger \sinh r, \\
    \hat{b}_3 &= \hat{b}_2 \cosh r - \hat{a}_2^\dagger \sinh r.
\label{eq:OPA2}
\end{split}
\end{equation}
Substituting Eqs.~(\ref{eqs:OPA1}) and (\ref{eq:displacement}) into Eq.~(\ref{eq:OPA2}) yields the output field operators immediately before detection:
\begin{equation}
\begin{split}
    \hat{a}_3 &= \hat{a}_0 + \beta \cosh r, \\
    \hat{b}_3 &= \hat{b}_0 - \beta^* \sinh r.
    \label{eq:output_operators}
\end{split}
\end{equation}
%




We now consider total intensity detection, described by the measurement operator
\begin{equation}
    \hat{M} = \hat{a}_3^\dagger\hat{a}_3+\hat{b}_3^\dagger \hat{b}_3,
\label{eq:signal}
\end{equation}
which corresponds to the sum of the output photon numbers in the two modes. For a two-mode vacuum input state $|0,0\rangle$, the expectation value and variance of $\hat{M}$ are
\begin{equation}
    \langle\hat{M}\rangle = \lvert \beta \rvert^2(\cosh^2r+\sinh^2r),
\label{eq:ideal_mean}
\end{equation}
\begin{equation}
    \Delta^2M = \lvert \beta \rvert^2(\cosh^2r+\sinh^2r).
\label{eq:ideal_var}
\end{equation}
The sensitivity for estimating the displacement magnitude $|\beta|$ is defined through the standard error-propagation formula,
\begin{equation}
    \Delta_{|\beta|} = \frac{\Delta \hat{M}}{ \partial _{|\beta|} \langle \hat{M} \rangle},
    \label{eq:sensitivity}
\end{equation}
where
\begin{equation}
\partial_{|\beta|}\langle\hat{M}\rangle
=2|\beta|\left(\cosh^2r+\sinh^2r\right).
\label{eq:partial}
\end{equation}
Substituting Eqs.~(\ref{eq:ideal_var}) and~(\ref{eq:partial}) into Eq.~(\ref{eq:sensitivity}) yields the ideal displacement sensitivity,
\begin{equation}
\Delta_{|\beta|}
=\frac{1}{2\sqrt{\cosh^2r+\sinh^2r}}.
\label{eq:sensitivity_beta}
\end{equation}

To verify the optimality of our proposed total intensity measurement scheme, we compare its sensitivity with the quantum Cramér--Rao bound (QCRB)~\cite{Braunstein_1994}. Following Ref.~\cite{Wang_2022}, the quantum Fisher information (QFI) for a two-mode coherent state $\ket{\gamma(\epsilon),\delta(\epsilon)}$, where $\epsilon$ denotes the parameter to be estimated, is given by
\begin{equation}
    \text{QFI}_{\epsilon}=4\left|\frac{\partial\gamma}{\partial\epsilon}\right|^2+4\left|\frac{\partial\delta}{\partial\epsilon}\right|^2.
\label{eq:QFI}
\end{equation}
In our scheme, the output state after the second OPA can be written as $\ket{\beta \cosh r, -\beta^* \sinh r}$. Substituting $\gamma = |\beta| e^{i\theta} \cosh r$ and $
\delta = -|\beta| e^{-i\theta} \sinh r$ together with $\epsilon = |\beta|$, yields
\begin{equation}
    \text{QFI}_{\lvert\beta\rvert}=4(\cosh^2r+\sinh^2r),
    \label{eq:10}
\end{equation}
from which the corresponding QCRB is
\begin{equation}
    \Delta^{\mathrm{QCRB}}_{ |\beta|} = \frac{1}{\sqrt{\mathrm{QFI}_{|\beta|}}}
    = \frac{1}{2\sqrt{\cosh^2 r + \sinh^2 r}}.
    \label{eq:QCRB_beta}
\end{equation}
Comparing Eqs.~\eqref{eq:sensitivity_beta} and \eqref{eq:QCRB_beta}, we find that the sensitivity of our proposed measurement scheme exactly saturates the QCRB for estimating the displacement magnitude $|\beta|$. This demonstrates that the total intensity measurement is quantum optimal under the ideal lossless condition. This scheme takes advantage of the strong photon number correlation between the output modes of the first OPA. The second OPA is necessary for the intensity measurement to achieve low measurement uncertainty, as discussed in Appendix. For the known-phase case,  the displacement can be estimated using a single-mode squeezed probe, with the squeezing phase satisfying $(\beta-2\theta=\pi)$~\cite{Agarwal2025}. The resulting sensitivity is enhanced by a factor of $\sqrt{2}$ relative to the QCRB of the phase-independent SU(1,1) scheme considered here.






\begin{figure}[bthp]
    \centering
    \includegraphics[width=0.48\textwidth]{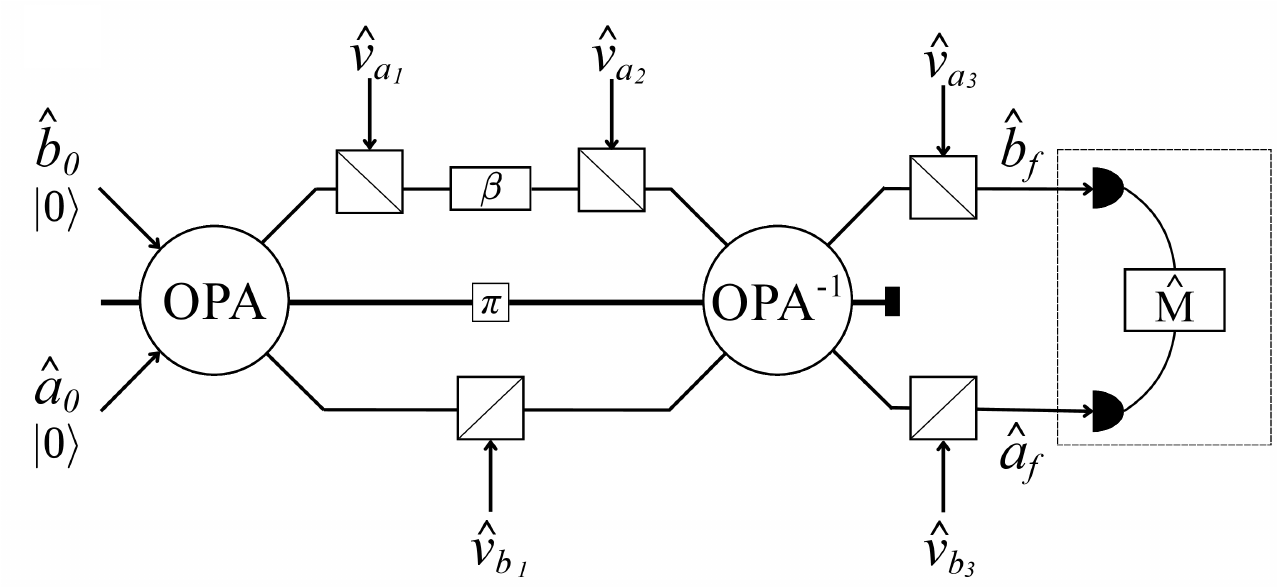}
    \caption{Schematics of the lossy SU(1,1) interferometer under two measurement schemes: Total intensity detection is performed on the two output modes, $\hat{a}_f$ and $\hat{b}_f$. Fictitious BSs with transmission coefficients $\eta_a$ and $\eta_b$ model photon loss in the two optical paths, while vacuum noise operators $\hat{v}{a_i}$ and $\hat{v}{b_i}$ enter through the open BS ports. The dark central line represents the strong classical pump field, which undergoes a $\pi$ phase shift between the two OPAs. $\hat{M}$ denotes the measurement operator acting on the detector outputs.}
    \label{fig:full_model}
\end{figure}

\renewcommand{\thesubsubsection}{\thesubsection.\arabic{subsubsection}}
\subsection{\label{sec:loss_model} Model with Loss}

We now consider a more realistic scenario by incorporating photon loss, following the standard beam splitter (BS) model described in Refs.~\cite{Marino_2012,Caves_2020}. In this model, fictitious BSs with transmission coefficients $\eta_a$ and $\eta_b$ couple vacuum fluctuations into the corresponding propagation modes, thereby accounting for optical losses. The output modes after transmission through the BSs are
\begin{equation}
    \begin{split}
    \hat{a}^{\prime}_i= \hat{a}_i\sqrt{\eta_a} + \hat{v}_{a_i}i\sqrt{1-\eta_a}, \\
    \hat{b}^{\prime}_i= \hat{b}_i\sqrt{\eta_b} + \hat{v}_{b_i}i\sqrt{1-\eta_b},
    \end{split}
    \label{loss_eqn}
\end{equation}
where $\hat{v}_{a_i}$ and $\hat{v}_{b_i}$ are the vacuum noise operators. 

The complete lossy SU(1,1) interferometer model is illustrated in Fig.~\ref{fig:full_model}. The corresponding output field operators after propagation through the lossy SU(1,1) interferometer are
\begin{equation}
\begin{aligned}
\hat{a}_f ={}&
i \hat{v}_{a_3}\sqrt{1-\eta_a} + i \hat{v}_{a_2}\sqrt{1-\eta_a}\sqrt{\eta_a}\cosh r \\
&+ i \hat{v}_{a_1}\sqrt{1-\eta_a}\,\eta_a \cosh r + e^{i\theta}\eta_a |\beta|\cosh r \\
&+ \hat{a}_0 \eta_a^{3/2}\cosh^2 r + i \hat{v}_{b_1}^\dagger\sqrt{\eta_a}\sqrt{1-\eta_b}\sinh r \\
&+ \hat{b}_0^\dagger \eta_a^{3/2}\cosh r \sinh r - \hat{b}_0^\dagger \sqrt{\eta_a}\sqrt{\eta_b}\cosh r \sinh r \\
&- \hat{a}_0 \sqrt{\eta_a}\sqrt{\eta_b}\sinh^2 r,
\end{aligned}
\label{a_final}
\end{equation}
\begin{equation}
\begin{aligned}
\hat{b}_f ={}&
i \hat{v}_{b_3}\sqrt{1-\eta_b} + i \hat{v}_{b_1}\sqrt{1-\eta_b}\sqrt{\eta_b}\cosh r \\
&+ \hat{b}_0 \eta_b \cosh^2 r + i \hat{v}_{a_2}^\dagger\sqrt{1-\eta_a}\sqrt{\eta_b}\sinh r \\
&+ i \hat{v}_{a_1}^\dagger\sqrt{1-\eta_a}\sqrt{\eta_a}\sqrt{\eta_b}\sinh r \\
&- e^{-i\theta}\sqrt{\eta_a}\sqrt{\eta_b}\,
|\beta|\sinh r - \hat{a}_0^\dagger \eta_a\sqrt{\eta_b}\cosh r \sinh r \\
&+ \hat{a}_0^\dagger \eta_b\cosh r \sinh r - \hat{b}_0 \eta_a\sqrt{\eta_b}\sinh^2 r.
\end{aligned}
\label{b_final}
\end{equation}

\subsubsection{\textbf{{\label{sec:full_model_QFI} 
Quantum Cram\'er--Rao Bound under Loss}}
}

The ultimate precision for the magnitude of displacement estimation achievable with the lossy SU(1,1) interferometer model shown in Fig.~\ref{fig:full_model} can be investigated using quantum Fisher information. 
The QFI measures how sensitively the quantum state changes under an infinitesimal change of a parameter. For a displacement with an unknown phase, the magnitude estimation is a multi-variable problem in general, where the relevant parameters are the magnitude ($|\beta|$) and the phase ($\theta$) of the displacement. However, for the lossy SU(1,1) interferometric model, we have found the off--diagonal terms in the QFI matrix to be zero, thus the phase provides no additional information about the magnitude of displacement. For a diagonal QFI matrix, the QCRB of $|\beta|$ estimation is found simply by taking the inverse of the respective element from the QFI matrix.

The two mode squeezed probe produced in the SU(1,1) interferometer is a state with a Gaussian Wigner function, and it remains Gaussian under the loss transformation shown in Eq.~\eqref{loss_eqn}~\cite{agarwal1987}. Thus, the quantum state can be fully specified using the displacement vector $(\textbf{d})$ and the covariance matrix $(\sigma)$. These quantities can be found using the transformation of the creation and annihilation operators in the Heisenberg picture. For a state undergoing transformations that preserve the Gaussian form, the QFI is given by~\cite{gao2014,safranek2019},
\begin{equation}
   \text{QFI}^{loss}_{|\beta|}=\lim_{\nu \to 1}\left[\tfrac{1}{2}\mathrm{vec}[\partial_i \sigma]^\dagger M^{-1}\mathrm{vec}[\partial_i \sigma]+2\partial_i \mathbf{d}^\dagger \sigma^{-1}\partial_i \mathbf{d}\right],
    \label{QFIgaussian}
\end{equation}
where $M=\left(\nu^2 \sigma \otimes \sigma - K \otimes K\right)$ and $K = \mathbb{I}_2 \oplus (-\mathbb{I}_2)$. Here $\mathbb{I}_2$ is the $2\times2$ identity matrix, and vec[.] denotes vectorization of a matrix.

We define the vector operator $\bm{\hat{\mathbf{A}}}_{\textbf{f}}\equiv(\hat a_{f},\hat b_{f},\hat a^\dagger_{f},\hat b^\dagger_{f})$ using the output field operators in Fig.~\ref{fig:full_model} as described in Eq.\eqref{a_final} and Eq.\eqref{b_final}. For a vacuum input state $(\rho_{in})$, the displacement vector can be obtained by taking the expectation value of $\bm{\hat{\mathbf{A}}}_{\textbf{f}}$,
\begin{equation}
    \begin{aligned}
    d^m=&\text{tr}(\hat\rho_{in}\hat {A}_{f}^m)\\
    =&\left(
\begin{array}{c}
 |\beta|  e^{i \theta } \eta _a \cosh (r) \\
 |\beta|  \left(-e^{-i \theta }\right) \sqrt{\eta _a} \sqrt{\eta _b} \sinh (r) \\
 |\beta|  e^{-i \theta } \eta _a \cosh (r) \\
 |\beta|  \left(-e^{i \theta }\right) \sqrt{\eta _a} \sqrt{\eta _b} \sinh (r) \\
\end{array}
\right)
    \end{aligned}
    \label{disp},
\end{equation}
Defining the fluctuation operator
$\Delta \hat A_f^m \equiv \hat A_f^m-d^m$, the covariance matrix is obtained from the symmetrized second central moments of the operator vector $\bm{\hat{\mathbf{A}}}_{f}$,
\begin{equation}
    \begin{aligned}
    \sigma^{mn}=\text{tr}(\hat\rho_{in}(\Delta\hat {A}_{f}^m\Delta\hat {A}_{f}^{n\dagger}+\Delta\hat {A}_{f}^n\Delta\hat {A}_{f}^{m\dagger}))
    \end{aligned}
    \label{covar}
\end{equation}
(Eqs.~(\ref{a_final}) and (\ref{b_final}) show that the dependence of the output field operators on the displacement parameter $\beta$ is through c-number terms. These terms are subtracted for the fluctuation operator $\Delta \hat{A}^{m}_{f}$. Therefore, the centered second moments are independent of $\beta$, and hence $\partial_{|\beta|}\sigma=0$. Consequently, the derivative of the covariance term in Eq.~\eqref{QFIgaussian} vanishes, and using Eq.~\eqref{disp} the QFI for estimating the displacement magnitude is given by,

\begin{equation}
    \begin{aligned}        \mathrm{QFI}^{\mathrm{loss}}_{|\beta|}&=2\,\partial_{|\beta|}\mathbf{d}^{\dagger}\sigma^{-1}\partial_{|\beta|}\mathbf{d}\\
    &=\frac{2-4P\sinh^2 r-Q\sinh^2(2r)}{8\beta^2\eta_a\left(\eta_a+R\sinh^2 r\right)},
    \label{QFI_loss_beta}
    \end{aligned}
    \end{equation}
    where
    \begin{align*}
    P &=
    -\eta_a-\eta_b
    +\eta_a(1+\eta_a)\eta_b
    \left[1+2(1-\eta_a)(1-\eta_b)\right],\\
    Q &=
    \left(\eta_a-\sqrt{\eta_b}\right)^2
    \left[\eta_a(2\eta_b-1)-\eta_b\right],\\
    R &=
    \eta_a+\eta_b
    -2\eta_a\eta_b(1-\eta_b).
\end{align*}
\begin{figure}[bthp]
    \centering
    \includegraphics[width=0.48\textwidth]{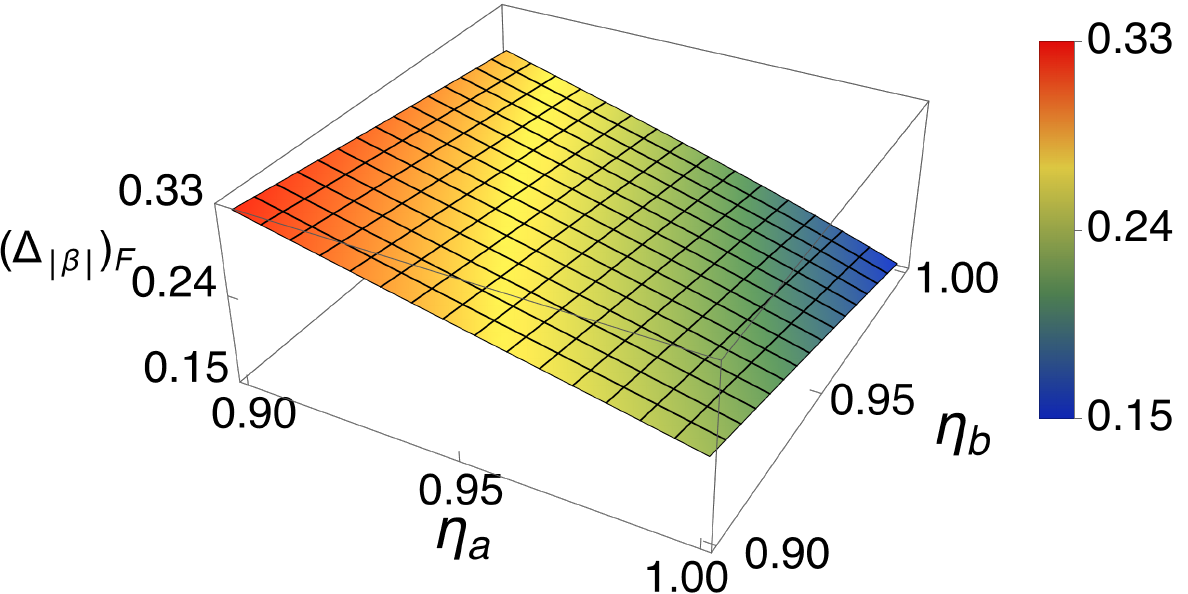}
    \caption{Minimum uncertainty for estimating the displacement magnitude, given by the quantum Cram\'er bound $(\Delta_{|\beta|})_F$ for a lossy SU(1,1) interferometer as a function of the transmission coefficients $\eta_a$ and $\eta_b$ in the high-transmission regime from 0.9 to 1, for a squeezing parameter of $r=1.5$.}
    \label{fig:QCRB}
\end{figure}
Taking the inverse of the QFI gives us the lowest uncertainty bound achievable for the displacement magnitude estimation with the lossy SU(1,1) interferometer given by the quantum Cram\'er--Rao  bound,
\begin{equation}
    \begin{aligned}
    (\Delta_{|\beta|})_{F}=&\frac{1}{\sqrt{\mathcal{N}\text{QFI}^{loss}_{|\beta|}}}\\
    =&\sqrt{\frac{
8\beta^2\eta_a
\left(\eta_a+R\sinh^2 r\right)
}{
2-4P\sinh^2 r-Q\sinh^2(2r)
}}
\end{aligned}
\end{equation},
where $\mathcal{N}$ is the number of independent measurements, which is taken to be one for simplicity. The QCRB $(\Delta_{|\beta|})_F$ is plotted in Fig.~\ref{fig:QCRB} as a function of the transmission coefficient $\eta_a$ and $\eta_b$ in the high transmission regime of 0.9 to 1, with the squeezing parameter $r$ fixed at 1.5. A smaller bound corresponds to a higher estimation precision. The bound increases monotonically as the transmission coefficients in either mode decreases. This increase is due to the uncorrelated noise introduced by the loss to both modes of the field. The stronger dependence on the loss in mode $\hat a$ is due to an additional loss channel and the attenuation of displacement. This increase in uncertainty bound represents a fundamental loss of information from the accessible output state. 

\subsubsection{\textbf{{\label{sec:full_model_intensity} 
Total Intensity Detection}}
}

\begin{figure*}[htbp]
    \centering
    \includegraphics[width=0.48\textwidth]{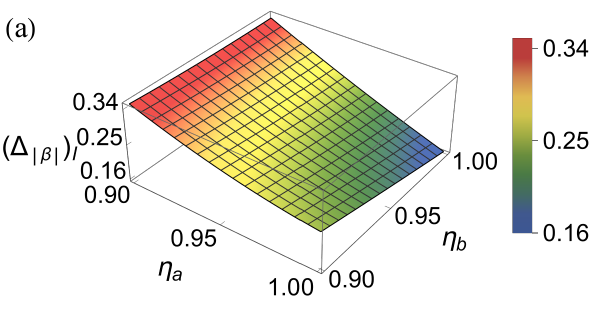}
    \hfill
    \includegraphics[width=0.48\textwidth]{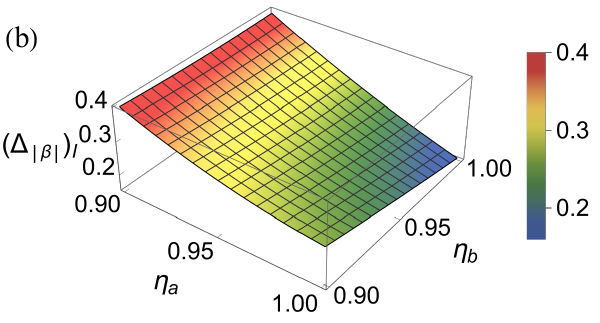}
    \caption{Displacement sensitivity $(\Delta_{|\beta|})_I$ of a lossy SU(1,1) interferometer employing the total intensity detection scheme as a function of the transmission coefficients $\eta_a$ and $\eta_b$ in the high-transmission regime from 0.9 to 1, for a squeezing parameter of $r=1.5$. Subfigures (a) and (b) correspond to coherent displacement amplitudes of $|\beta|=1$ and $|\beta|=1/2$, respectively, yielding $\mathrm{SNR}=1$ and $\mathrm{SNR}=1/2$. These cases represent operating regimes in which the mean coherent displacement is at or below the uncertainty associated with vacuum fluctuations.}
    \label{fig:intensity_S_high_trans}
    
\end{figure*}
\begin{figure*}[bthp]
    \centering \includegraphics[width=0.48\textwidth]{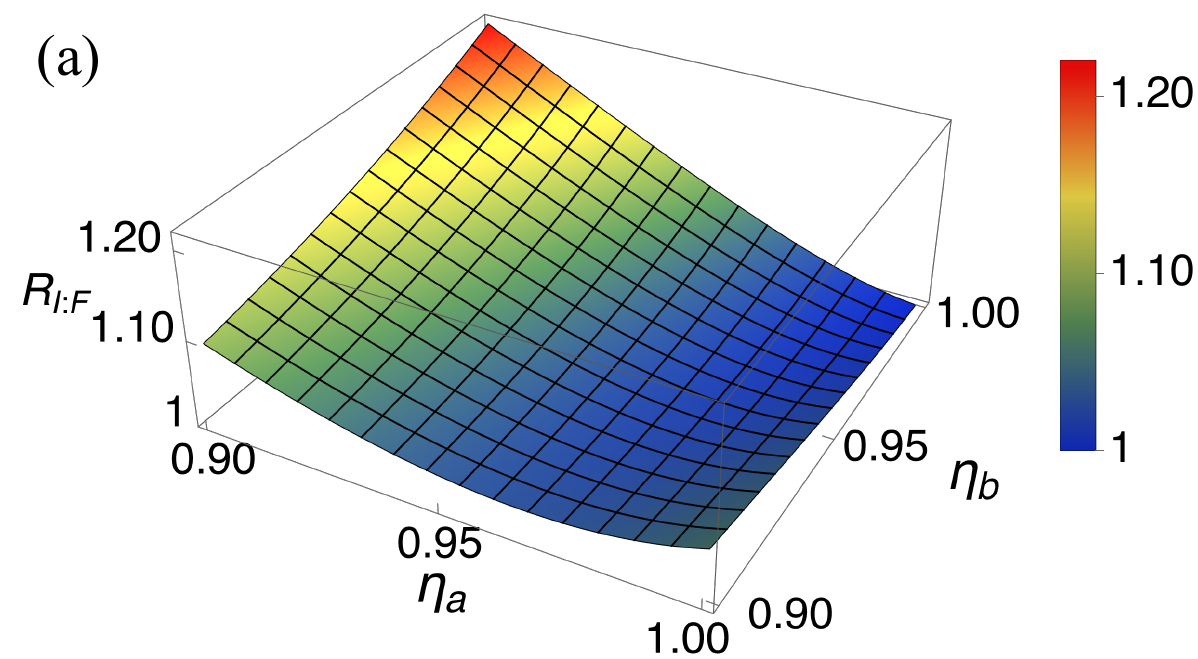}
    \hfill
    \includegraphics[width=0.48\textwidth]{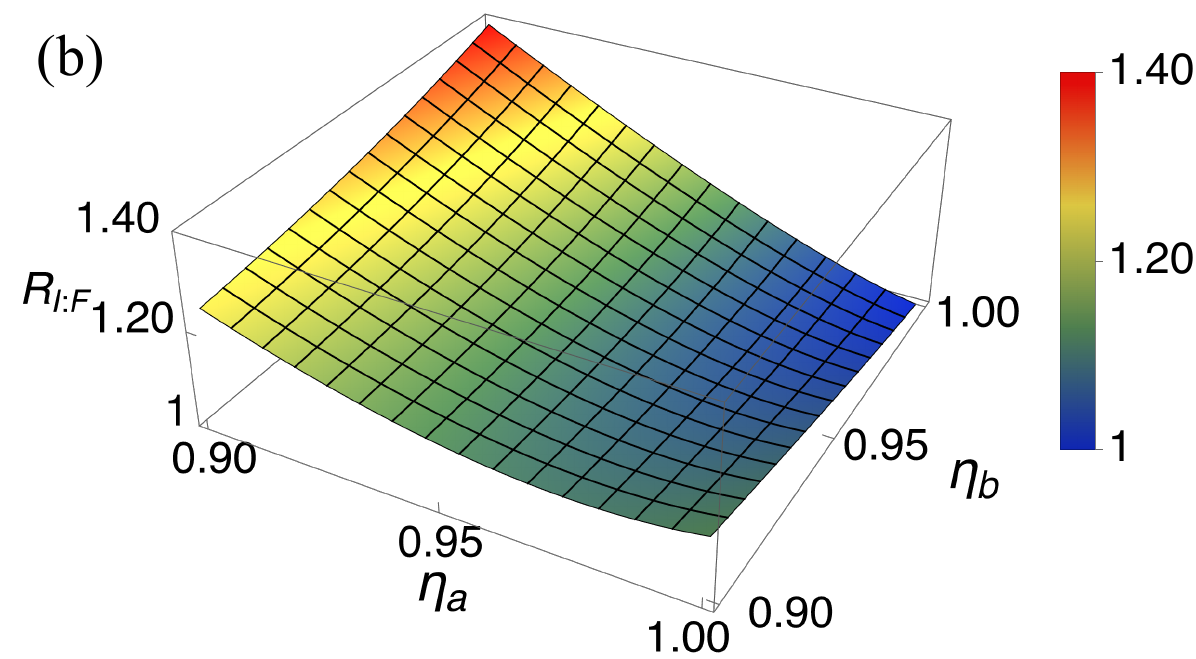}
    \caption{Ratio between the sensitivity with total intensity detection and the quantum Cram\'er--Rao bound, $(\Delta_{|\beta|})_I/(\Delta_{|\beta|})_F$, for displacement-magnitude estimation in the lossy SU(1,1) interferometer as a function of transmission coefficients $\eta_a \text{ and } \eta_b$ in the high-transmission regime, $0.9\leq\eta_a,\eta_b\leq1$. The squeezing parameter is fixed at 1.5. Subfigures (a) and (b) correspond to coherent displacement amplitudes of $|\beta|=1$ and $|\beta|=1/2$, respectively, yielding $\mathrm{SNR}=1$ and $\mathrm{SNR}=1/2$.}
\label{fig:S_QFI_ratio}
    
\end{figure*}

The measurement operator for the lossy SU(1,1) interferometer in Fig.~\ref{fig:full_model} is again defined as
\begin{equation}
    \hat{M_I} = \hat{a}_f^\dagger\hat{a}_f+\hat{b}_f^\dagger \hat{b}_f.
\label{eq:SI_signal}
\end{equation}
Evaluating the expectation value with respect to the input two-mode vacuum state $|0,0\rangle$ and all auxiliary vacuum modes introduced by the BSs, we find
\begin{equation}
\begin{aligned}
\langle \hat{M}_{I} \rangle =
{}&|\beta|^2\,\eta_a\,
\Bigl(
\eta_a \cosh^2 r
+ \eta_b \sinh^2 r
\Bigr) \\
&+ \Bigl(
\eta_a + \eta_b
- \eta_a (1+\eta_a)\eta_b \\
&
+ (\eta_a - \sqrt{\eta_b})^2
(\eta_a + \eta_b)\cosh^2 r
\Bigr)\sinh^2 r. \\
\end{aligned}
\label{eq:SI_mean}
\end{equation}
The first term is proportional to $|\beta|^2$ and constitutes the displacement-dependent signal, while the remaining terms represent a loss-induced background independent of $|\beta|$. In the lossless limit $\eta_a = \eta_b = 1$, the background vanishes and Eq.~\eqref{eq:SI_mean} reduces to Eq.~\eqref{eq:ideal_mean} in the ideal lossless model.

Due to the lengthy analytical expressions of the variance $\Delta^2M_I$ and sensitivity $(\Delta_{|\beta|})_I$, the full expressions of them are intentionally omitted. In Fig.~\ref{fig:intensity_S_high_trans}, we plot the sensitivity $(\Delta_{|\beta|})_I$ as a function of the transmission coefficients $\eta_a$ and $\eta_b$ with the squeezing parameter fixed at $r=1.5$. Figures~\ref{fig:intensity_S_high_trans}(a) and \ref{fig:intensity_S_high_trans}(b) correspond to coherent displacement amplitudes of $|\beta|=1$ and $|\beta|=1/2$, respectively, corresponding to signal-to-noise ratios (SNRs) of 1 and 1/2. These two cases represent operating regimes in which \textit{the mean coherent displacement is at or below the uncertainty associated with vacuum fluctuations.}

Fig.~\ref{fig:S_QFI_ratio} compares the sensitivity of total intensity detection with the quantum Cram\'er Rao bound by plotting the ratio $(\Delta_{|\beta|})_I/(\Delta_{|\beta|})_F$ as a function of the transmission coefficients. A value larger than unity denotes the excess uncertainty in the measurement relative to the uncertainty bound allowed by the quantum state. In the high transmission regime the ratio remains close to unity showing that for low optical loss the total intensity measurement is nearly optimal. As either transmission coefficient decreases, the ratio increases, indicating that the sensitivity of the measurement becomes worse relative to the QCRB due to the additional noise introduced by the loss. Although total intensity measurement is not optimal in general, for the whole plotted region the deviation of sensitivity from the QCRB remains below $22\%$ and $39\%$ for displacement of magnitude $|\beta|=1 \text{ and }|\beta|=1/2$, as shown in Figs.~\ref{fig:S_QFI_ratio}(a) and~\ref{fig:S_QFI_ratio}(b), respectively. The larger deviation for $|\beta|=1/2$ is expected because for a smaller displacement the intensity signal is smaller relative to the additional noise introduced by loss. Despite this deviation, the total intensity detection remains a practical phase-independent displacement magnitude measurement scheme in the relevant high-transmission regime.

%
    

%
\begin{figure}[bt]
    \centering
    \includegraphics[width=\columnwidth]{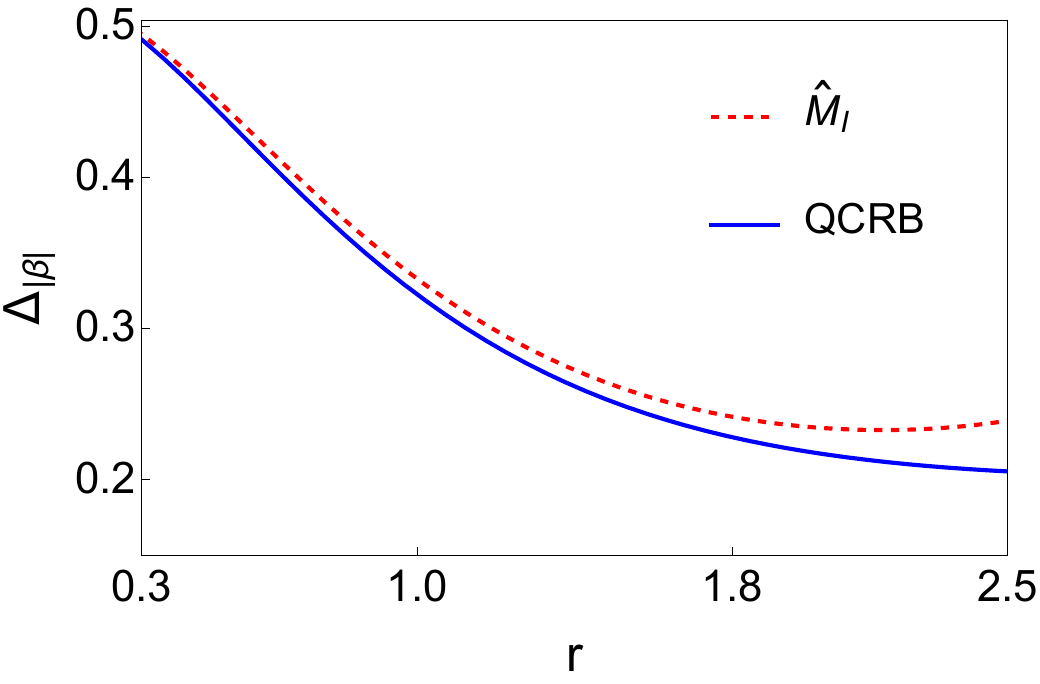}
    \caption{Displacement sensitivity and the QCRB, $\Delta_{|B|}$, as a function of the squeezing parameter $r$ for the measurement scheme considered, assuming $\eta_a=\eta_b=0.95$ and $|\beta|=1$:\ the QCRB (blue solid) and the total intensity detection ($\hat{M}_I$, red dotted) with the conventional SU(1,1).}
    \label{fig:all_signals_varying_r}
\end{figure}

We now compare the sensitivity for the SU(1,1)-based total intensity measurement ($\hat M_I$) with the lowest uncertainty bound attainable with the SU(1,1) scheme set by the QCRB, as a function of the squeezing parameter r, assuming $\eta_a=\eta_b=0.95 \text{ and } |\beta|=1$. The results are shown in Fig.~\ref{fig:all_signals_varying_r}. The plot shows that the QCRB improves with an increase in the squeeze parameter. This is because for larger squeezing, there are more correlated photons available for the estimation of displacement. With total intensity measurement, the sensitivity approaches the QCRB in the range of $r<2.18$, reaching the minimum at $2.18$. At larger values of squeezing, there is a large intensity fluctuation which worsens the sensitivity of the measurement. For the entire range of the plot, the measurement sensitivity remains close to the QCRB, showing near optimality, with a maximum deviation of only $16\%$.

\section{\label{sec:results}Conclusion}

In this work, we have developed a comprehensive theoretical framework for phase-independent detection of weak coherent optical displacements based on SU(1,1) interferometry. We showed that, under ideal lossless conditions, a conventional SU(1,1) interferometer employing only total intensity detection saturates the QCRB for estimating the displacement magnitude, demonstrating that coherent homodyne detection is not fundamentally required for quantum-optimal phase-independent sensing. 

We further derived the analytical expression of QCRB and the sensitivity of the conventional SU(1,1) interferometer with total intensity detection and systematically investigated their performance in the presence of optical loss. Our results show that, when the displacement phase is unknown, total-intensity detection with an SU(1,1) interferometer achieves an estimation uncertainty close to the quantum Cramér–Rao bound, demonstrating near-optimal displacement-magnitude sensitivity. In this scheme, the absence of off--diagonal terms in the QFI matrix shows that the knowledge of phase provides no additional information about the displacement magnitude. Thus, the absence of phase information does not impose any limitation on this scheme. These results establish SU(1,1)-based intensity detection as a practical platform for phase-independent quantum sensing where the signal phase is unknown, fluctuating, or otherwise inaccessible.

\section*{Acknowledgments} 
LC, GSA, and TL acknowledge support from the U.S. National Science Foundation (NSF) through the ExpandQISE program under Award No. 2426699. LC and TL acknowledge support from the NSF CCSS program under Award No. 2503630. PD, HPL, and TL also acknowledge support from the U.S. National Institute of Standards and Technology (NIST) through the CIPP program under Award No. 60NANB24D218.

\appendix
\setcounter{equation}{0}
\renewcommand{\theequation}{A\arabic{equation}}
\section*{Appendix: Sensitivity of the Single-OPA SU(1,1) Scheme}
\begin{figure}[tpbh]
\centering
\includegraphics[width=\columnwidth]{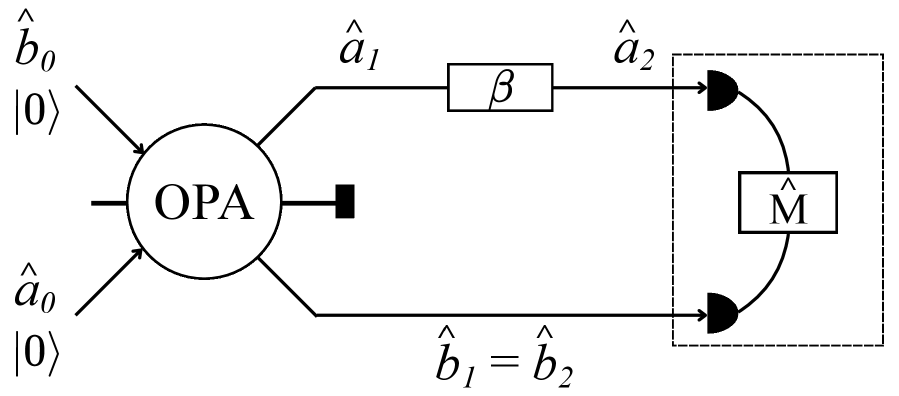}
\caption{Schematic of balanced detection measurement scheme based on an idealized lossless SU(1,1) interferometer with a single OPA. Input modes $\hat{a}_{0}$ and $\hat{b}_{0}$ are injected into the first OPA, generating correlated output modes $\hat{a}_{1}$ and $\hat{b}_{1}$. A weak coherent displacement operation $\hat{D}(\beta)$ is subsequently applied to mode $\hat{a}_{1}$. The central dark line denotes a strong classical pump field. The balanced detection measurement operator $\hat{M}$ is performed by detecting the output intensity difference of modes $\hat{a}_{2}$ and $\hat{b}_{2}$.}
\label{fig:balanced_detection_schematics}
\end{figure}

In this Appendix we analyze the ideal lossless SU(1,1) scheme without the second OPA and compare it with the two-OPA scheme shown in Fig.~\ref{fig:toy_model}, to highlight the importance of the second OPA. Fig.~\ref{fig:balanced_detection_schematics} shows the corresponding SU(1,1) scheme with a single OPA. In this scheme, the difference in photon number between the two modes is measured using a balanced detection. The balanced detection measurement operator is

\begin{equation}
    \hat{M_b} = \hat{a}_2^\dagger\hat{a}_2-\hat{b}_2^\dagger \hat{b}_2,
\label{eq:BD_signal}
\end{equation}
which corresponds to the difference of the output photon numbers between the two modes. For a two-mode vacuum input state $|0,0\rangle$, the expectation value and variance of $\hat{M_b}$ can be found using Eqs.~\eqref{eqs:OPA1} and~\eqref{eq:displacement},
\begin{equation}
    \langle\hat{M_b}\rangle = \lvert \beta \lvert^2,
\label{eq:BD_mean}
\end{equation}
\begin{equation}
    \Delta^2M_b =\lvert \beta \rvert^2 (\sinh^2 r + \cosh^2 r).
\label{eq:BD_var}
\end{equation}

 Following steps similar to section~\ref{sec:ideal_model}, we can find the sensitivity in this case.
The sensitivity for estimating the displacement magnitude $|\beta|$ is defined through the standard error-propagation formula,
\begin{equation}
    \Delta_{|\beta|} = \frac{\Delta \hat{M_b}}{ \partial _{|\beta|} \langle \hat{M_b} \rangle},
    \label{eq:BD_sensitivity}
\end{equation}
where
\begin{equation}
\partial_{|\beta|}\langle\hat{M_b}\rangle
=2|\beta|.
\label{eq:BD_partial}
\end{equation}
Substituting Eqs.~(\ref{eq:BD_var}) and~(\ref{eq:BD_partial}) into Eq.~(\ref{eq:BD_sensitivity}) yields the ideal displacement sensitivity,
\begin{equation}
\Delta_{|\beta|}
=\frac{1}{4} \cosh (2 r).
\label{eq:BDsensitivity_beta}
\end{equation}

Comparing this result with Eq.~\eqref{eq:sensitivity_beta}, we find that the error sensitivity with the single-OPA scheme becomes larger as the squeezing parameter increases. This occurs because stronger squeezing increases the noise without increasing the mean signal. As a result, intensity-difference detection with a single OPA performs worse than a vacuum probe. Thus, although the entangled source in an SU(1,1) interferometer provides a large quantum Fisher information, achieving high sensitivity also requires a suitable measurement, where the signal must vary strongly with the estimated parameter while the noise remains low. The total intensity measurement with the second OPA provides this advantage by amplifying the parameter-dependent signal as the squeezing increases. Meanwhile, comparison of Eqs.~\eqref{eq:BD_var} and \eqref{eq:ideal_var} shows that the noise remains the same as in the single-OPA scheme, including its dependence on squeezing. Thus, the second OPA amplifies the signal without adding further noise. 

\bibliographystyle{apsrev4-2}
\bibliography{reference}

@article{Agarwal2025,
  title = {Saturation of the quantum Cram\'er-Rao bound for distributed sensing via error sensitivity in SU(1,1)-SU($m$) interferometry},
  author = {Agarwal, Girish S.},
  journal = {Phys. Rev. A},
  volume = {112},
  issue = {3},
  pages = {032439},
  numpages = {9},
  year = {2025},
  month = {Sep},
  publisher = {American Physical Society},
  doi = {10.1103/h2x6-dz96},
  url = {https://link.aps.org/doi/10.1103/h2x6-dz96}
}

@article{Wang_2022,
   title={Quantum Fisher information perspective on sensing in anti-PT symmetric systems},
   volume={4},
   ISSN={2643-1564},
   url={http://dx.doi.org/10.1103/PhysRevResearch.4.013131},
   DOI={10.1103/physrevresearch.4.013131},
   number={1},
   journal={Physical Review Research},
   publisher={American Physical Society (APS)},
   author={Wang, J. and Mukhopadhyay, D. and Agarwal, G. S.},
   year={2022},
   month=feb }

@article{Yurke_1986,
  title = {SU(2) and SU(1,1) interferometers},
  author = {Yurke, Bernard and McCall, Samuel L. and Klauder, John R.},
  journal = {Phys. Rev. A},
  volume = {33},
  issue = {6},
  pages = {4033--4054},
  numpages = {0},
  year = {1986},
  month = {Jun},
  publisher = {American Physical Society},
  doi = {10.1103/PhysRevA.33.4033},
  url = {https://link.aps.org/doi/10.1103/PhysRevA.33.4033}
}

@article{Braunstein_1994,
  title = {Statistical distance and the geometry of quantum states},
  author = {Braunstein, Samuel L. and Caves, Carlton M.},
  journal = {Phys. Rev. Lett.},
  volume = {72},
  issue = {22},
  pages = {3439--3443},
  numpages = {0},
  year = {1994},
  month = {May},
  publisher = {American Physical Society},
  doi = {10.1103/PhysRevLett.72.3439},
  url = {https://link.aps.org/doi/10.1103/PhysRevLett.72.3439}
}

@article{Caves_2020,
   title={Reframing SU(1,1) Interferometry},
   volume={3},
   ISSN={2511-9044},
   url={http://dx.doi.org/10.1002/qute.201900138},
   DOI={10.1002/qute.201900138},
   number={11},
   journal={Advanced Quantum Technologies},
   publisher={Wiley},
   author={Caves, Carlton M.},
   year={2020},
   month=Mar }

@article{Ou_2020,
       author = {{Ou}, Z.~Y. and {Li}, Xiaoying},
        title = "{Quantum SU(1,1) interferometers: Basic principles and applications}",
      journal = {APL Photonics},
         year = 2020,
        month = aug,
       volume = {5},
       number = {8},
          eid = {080902},
        pages = {080902},
          doi = {10.1063/5.0004873},
archivePrefix = {arXiv},
       eprint = {2004.12469},
 primaryClass = {quant-ph},
       adsurl = {https://ui.adsabs.harvard.edu/abs/2020APLP....5h0902O}
}

@article{Ou_2012,
  title = {Enhancement of the phase-measurement sensitivity beyond the standard quantum limit by a nonlinear interferometer},
  author = {Ou, Z. Y.},
  journal = {Phys. Rev. A},
  volume = {85},
  issue = {2},
  pages = {023815},
  numpages = {7},
  year = {2012},
  month = {Feb},
  publisher = {American Physical Society},
  doi = {10.1103/PhysRevA.85.023815},
  url = {https://link.aps.org/doi/10.1103/PhysRevA.85.023815}
}

@book{gerry2005introductory,
  title = {Introductory Quantum Optics},
  author = {Gerry, Christopher C. and Knight, Peter L.},
  year = {2005},
  publisher = {Cambridge University Press},
  address = {Cambridge, UK},
  isbn = {9780521820356}
}

@article{Marino_2012,
  title = {Effect of losses on the performance of an SU(1,1) interferometer},
  author = {Marino, A. M. and Corzo Trejo, N. V. and Lett, P. D.},
  journal = {Phys. Rev. A},
  volume = {86},
  issue = {2},
  pages = {023844},
  numpages = {8},
  year = {2012},
  month = {Aug},
  publisher = {American Physical Society},
  doi = {10.1103/PhysRevA.86.023844},
  url = {https://link.aps.org/doi/10.1103/PhysRevA.86.023844}
}

@article{Anderson_2017,
  title = {Optimal phase measurements with bright- and vacuum-seeded SU(1,1) interferometers},
  author = {Anderson, Brian E. and Schmittberger, Bonnie L. and Gupta, Prasoon and Jones, Kevin M. and Lett, Paul D.},
  journal = {Phys. Rev. A},
  volume = {95},
  issue = {6},
  pages = {063843},
  numpages = {9},
  year = {2017},
  month = {Jun},
  publisher = {American Physical Society},
  doi = {10.1103/PhysRevA.95.063843},
  url = {https://link.aps.org/doi/10.1103/PhysRevA.95.063843}
}

@article{LIGO_2013,
   title={Enhanced sensitivity of the LIGO gravitational wave detector by using squeezed states of light},
   volume={7},
   ISSN={1749-4893},
   url={http://dx.doi.org/10.1038/nphoton.2013.177},
   DOI={10.1038/nphoton.2013.177},
   number={8},
   journal={Nature Photonics},
   publisher={Springer Science and Business Media LLC},
   author={Aasi, J. and Abadie, J. and Abbott, B. P. and others},
   year={2013},
   month=July, 
   pages={613–619} }

@article{LIGO_2019,
  title = {Quantum-Enhanced Advanced LIGO Detectors in the Era of Gravitational-Wave Astronomy},
  author = {Tse, M. and Yu, Haocun and Kijbunchoo, N. and others},
  journal = {Phys. Rev. Lett.},
  volume = {123},
  issue = {23},
  pages = {231107},
  numpages = {8},
  year = {2019},
  month = {Dec},
  publisher = {American Physical Society},
  doi = {10.1103/PhysRevLett.123.231107},
  url = {https://link.aps.org/doi/10.1103/PhysRevLett.123.231107}
}

@article{VIRGO_2019,
  title = {Increasing the Astrophysical Reach of the Advanced Virgo Detector via the Application of Squeezed Vacuum States of Light},
  author = {Acernese, F. and Agathos, M. and Aiello, L. and others},
  collaboration = {Virgo Collaboration},
  journal = {Phys. Rev. Lett.},
  volume = {123},
  issue = {23},
  pages = {231108},
  numpages = {10},
  year = {2019},
  month = {Dec},
  publisher = {American Physical Society},
  doi = {10.1103/PhysRevLett.123.231108},
  url = {https://link.aps.org/doi/10.1103/PhysRevLett.123.231108}
}

@article{Backes_2021,
   title={A quantum enhanced search for dark matter axions},
   volume={590},
   ISSN={1476-4687},
   url={http://dx.doi.org/10.1038/s41586-021-03226-7},
   DOI={10.1038/s41586-021-03226-7},
   number={7845},
   journal={Nature},
   publisher={Springer Science and Business Media LLC},
   author={Backes, K. M. and Palken, D. A. and Kenany, S. Al and others},
   year={2021},
   month=Feb, pages={238–242} }

@misc{Zheng_2016,
      title={Accelerating dark-matter axion searches with quantum measurement technology}, 
      author={Huaixiu Zheng and Matti Silveri and R. T. Brierley and S. M. Girvin and K. W. Lehnert},
      year={2016},
      eprint={1607.02529},
      archivePrefix={arXiv},
      primaryClass={hep-ph},
      url={https://arxiv.org/abs/1607.02529}, 
}

@article{D_Li_2014,
   title={The phase sensitivity of an SU(1,1) interferometer with coherent and squeezed-vacuum light},
   volume={16},
   ISSN={1367-2630},
   url={http://dx.doi.org/10.1088/1367-2630/16/7/073020},
   DOI={10.1088/1367-2630/16/7/073020},
   number={7},
   journal={New Journal of Physics},
   publisher={IOP Publishing},
   author={Li, Dong and Yuan, Chun-Hua and Ou, Z Y and Zhang, Weiping},
   year={2014},
   month=July, pages={073020} }

@article{Genoni_2013,
  title = {Optimal estimation of joint parameters in phase space},
  author = {Genoni, M. G. and Paris, M. G. A. and Adesso, G. and Nha, H. and Knight, P. L. and Kim, M. S.},
  journal = {Phys. Rev. A},
  volume = {87},
  issue = {1},
  pages = {012107},
  numpages = {7},
  year = {2013},
  month = {Jan},
  publisher = {American Physical Society},
  doi = {10.1103/PhysRevA.87.012107},
  url = {https://link.aps.org/doi/10.1103/PhysRevA.87.012107}
}

@article{Hanamura_2023,
  title = {Single-Shot Single-Mode Optical Two-Parameter Displacement Estimation beyond Classical Limit},
  author = {Hanamura, Fumiya and Asavanant, Warit and Kikura, Seigo and Mishima, Moeto and Miki, Shigehito and Terai, Hirotaka and Yabuno, Masahiro and China, Fumihiro and Fukui, Kosuke and Endo, Mamoru and Furusawa, Akira},
  journal = {Phys. Rev. Lett.},
  volume = {131},
  issue = {23},
  pages = {230801},
  numpages = {5},
  year = {2023},
  month = {Dec},
  publisher = {American Physical Society},
  doi = {10.1103/PhysRevLett.131.230801},
  url = {https://link.aps.org/doi/10.1103/PhysRevLett.131.230801}
}

@article{He_2023,
   title={Quantum microscopy of cells at the Heisenberg limit},
   volume={14},
   ISSN={2041-1723},
   url={http://dx.doi.org/10.1038/s41467-023-38191-4},
   DOI={10.1038/s41467-023-38191-4},
   number={1},
   journal={Nature Communications},
   publisher={Springer Science and Business Media LLC},
   author={He, Zhe and Zhang, Yide and Tong, Xin and Li, Lei and Wang, Lihong V.},
   year={2023},
   month=Apr }

@article{Taylor_2013,
   title={Biological measurement beyond the quantum limit},
   volume={7},
   ISSN={1749-4893},
   url={http://dx.doi.org/10.1038/nphoton.2012.346},
   DOI={10.1038/nphoton.2012.346},
   number={3},
   journal={Nature Photonics},
   publisher={Springer Science and Business Media LLC},
   author={Taylor, Michael A. and Janousek, Jiri and Daria, Vincent and Knittel, Joachim and Hage, Boris and Bachor, Hans-A. and Bowen, Warwick P.},
   year={2013},
   month=Feb, pages={229–233} }

@article{Yokoyama_2013,
   title={Ultra-large-scale continuous-variable cluster states multiplexed in the time domain},
   volume={7},
   ISSN={1749-4893},
   url={http://dx.doi.org/10.1038/nphoton.2013.287},
   DOI={10.1038/nphoton.2013.287},
   number={12},
   journal={Nature Photonics},
   publisher={Springer Science and Business Media LLC},
   author={Yokoyama, Shota and Ukai, Ryuji and Armstrong, Seiji C. and Sornphiphatphong, Chanond and Kaji, Toshiyuki and Suzuki, Shigenari and Yoshikawa, Jun-ichi and Yonezawa, Hidehiro and Menicucci, Nicolas C. and Furusawa, Akira},
   year={2013},
   month=Nov, pages={982–986} }

@article{Furusawa_1998,
author = {A. Furusawa  and J. L. Sørensen  and S. L. Braunstein  and C. A. Fuchs  and H. J. Kimble  and E. S. Polzik },
title = {Unconditional Quantum Teleportation},
journal = {Science},
volume = {282},
number = {5389},
pages = {706-709},
year = {1998},
doi = {10.1126/science.282.5389.706},
URL = {https://www.science.org/doi/abs/10.1126/science.282.5389.706}}

@article{Lopaeva_2013,
   title={Experimental Realization of Quantum Illumination},
   volume={110},
   ISSN={1079-7114},
   url={http://dx.doi.org/10.1103/PhysRevLett.110.153603},
   DOI={10.1103/physrevlett.110.153603},
   number={15},
   journal={Physical Review Letters},
   publisher={American Physical Society (APS)},
   author={Lopaeva, E. D. and Ruo Berchera, I. and Degiovanni, I. P. and Olivares, S. and Brida, G. and Genovese, M.},
   year={2013},
   month=Apr }

@article{Gregory_2020,
author = {T. Gregory  and P.-A. Moreau  and E. Toninelli  and M. J. Padgett },
title = {Imaging through noise with quantum illumination},
journal = {Science Advances},
volume = {6},
number = {6},
pages = {eaay2652},
year = {2020},
doi = {10.1126/sciadv.aay2652},
URL = {https://www.science.org/doi/abs/10.1126/sciadv.aay2652}}

@article{Blakey_2022,
   title={Quantum and non-local effects offer over 40 dB noise resilience advantage towards quantum lidar},
   volume={13},
   ISSN={2041-1723},
   url={http://dx.doi.org/10.1038/s41467-022-33376-9},
   DOI={10.1038/s41467-022-33376-9},
   number={1},
   journal={Nature Communications},
   publisher={Springer Science and Business Media LLC},
   author={Blakey, Phillip S. and Liu, Han and Papangelakis, Georgios and Zhang, Yutian and Léger, Zacharie M. and Iu, Meng Lon and Helmy, Amr S.},
   year={2022},
   month=Sept }

@software{NCAlgebra,
  author = {J. William Helton and Mauricio C. de Oliveira},
  title = {{NCAlgebra: Non Commutative Algebra Package for Mathematica}},
  url = {https://github.com/NCAlgebra/NC},
  version = {6.0.3},
  year = {2023}
}

@article{Agarwal1987,
author = {G.S. Agarwal},
title = {Wigner-function Description of Quantum Noise in Interferometers},
journal = {Journal of Modern Optics},
volume = {34},
number = {6-7},
pages = {909--921},
year = {1987},
publisher = {Taylor \& Francis},
doi = {10.1080/09500348714550831},
}

@article{gao2014,
  title={Bounds on quantum multiple-parameter estimation with Gaussian state},
  author={Gao, Yang and Lee, Hwang},
  journal={The European Physical Journal D},
  volume={68},
  number={11},
  pages={347},
  year={2014},
  publisher={Springer}
}

@article{safranek2019,
doi = {10.1088/1751-8121/aaf068},
url = {https://doi.org/10.1088/1751-8121/aaf068},
year = {2018},
month = {dec},
publisher = {IOP Publishing},
volume = {52},
number = {3},
pages = {035304},
author = {Šafránek, Dominik},
title = {Estimation of Gaussian quantum states},
journal = {Journal of Physics A: Mathematical and Theoretical}
}

\end{document}